# Hispidin and Lepidine E: two Natural Compounds and Folic acid as Potential Inhibitors of 2019-novel coronavirus Main Protease (2019-nCoVM$^{pro}$), molecular docking and SAR study

Talia SERSEG[1*], Khedidja BENAROUS[1] & Mohamed YOUSFI[1]

[1]Fundamental sciences laboratory, Amar Telidji University, Laghouat, Algeria

**Abstract**

2019-nCoV is a novel coronavirus was isolated and identified in 2019 in Wuhan, China. On 17th February and according to world health organization, a number of 71 429 confirmed cases worldwide, among them 2162 new cases recorded in the last 24 hours. There is no drug or vaccine for human and animal coronavirus. The inhibition of 3CL hydrolase enzyme provides a promising therapeutic principle for developing treatments against CoViD-19. The 3CLpro (Mpro) known for involving in counteracting the host innate immune response. This work presents the inhibitory effect of some natural compounds against 3CL hydrolase enzyme, and explain the main interactions in inhibitor-enzyme complex. Molecular docking study carried out using Autodock Vina. By screening several molecules, we identified three candidate agents that inhibit the main protease of coronavirus. Hispidin, lepidine E, and folic acid bound tightly in the enzyme, strong hydrogen bonds have been formed (1.69-1.80Å) with the active site residues. This study provides a possible therapeutic strategy for CoViD-19.

## 1. Introduction

Coronaviruses include a large and diverse family of enveloped, positive-stranded RNA viruses [1]. The latest International Committee of Taxonomy of Viruses (ICTV) classification indicate that the thirty-eight species of coronaviruses belong to one of four genera (α-, β-, δ-, and γ-) [2], where α- and β-CoVs principally infect mammals, seventeen species among them hosted in bats (BtCoV-HKU10, BtCoV-CDPHE15, BtRfCoV-HuB13, BtMiCoV-1, BtMiCoV-HKU8, BtMy-Sax11, BtNy-Sc13, BtScCoV-512, BtRhCoV-HKU2, BtKYNL63, BtHpCoV-ZJ13, MERSr-CoV, BtPiCoV-HKU5, BtTyCoV-HKU4, BtEoCoV-GCCDC1, BtRoCoV-HKU9 and SARSr-CoV), γ- and δ-CoVs principally infect birds. Six species of CoVs: α-CoVs 229E and NL63 and β-CoVs HCoV-HKU1, MERSr-CoV, SARSr-CoV and HCoV-OC43 can infect human causing diseases. [3-6]. SARS-CoV and MERS-CoV were the causal agents of respiratory disease outbreaks in 2002 and 2012 in Guangdong Province, China, and the Middle East, respectively [4,5,7,8]. A novel coronavirus was isolated and identified in 2019 in Wuhan, China, which was named 2019 novel coronavirus (2019-nCoV) [8,9].

----

*Address correspondence to this author at the Department of Biology, Faculty of science, Amar TELIDJI university, Route de Ghardaïa, BP 37G, 03000 Laghouat, Algérie; E-mails: t.serseg@ens-lagh.dz, k.benarous@lagh-univ.dz; yousfim8@gmail.com

On 17th February and according to world health organization, a number of 71 429 confirmed cases worldwide, among them 2162 new cases recorded in last 24 hours. 2019-nCoV caused 1772 deaths in china and 3 death outside [10]. The coronavirus 3CL hydrolase (Mpro) enzyme, also known as the main protease, exhibits an important function in the viral life-cycle. Its inhibition provides a promising therapeutic principle for developing treatments against CoViD-19 to prevent further spread [11-13].

The 3CL hydrolase (Mpro) is one of coronaviruses' proteases. CoVMpro has become an attractive target for anti-CoV drug design, due to his responsibility for the maturation of itself and key functional enzymes such as:replicase and helicase. It is involved in counteracting the host innate immune response [14-16]. It has been reported that The Mpros of two CoVs infecting pigs antagonize the host immune response. They cleave porcine IKKγ (NEMO) at the identical site, Gln231↓Val232, where NEMO is required for activating the NF-κB and IRF3 pathways, which abrogates NF-κB signaling and inhibits IFNβ induction [17,18]. the Mpro of PDCoV also impair the JAK-STAT pathway by processing porcine STAT2 at two sites, Gln685↓Glu686 and Gln758↓Ser759 [14,19].

Hispidin (6-(3,4-dihydroxystyryl)-4-hydroxy-2-pyrone) is a polyphenol and an important medicinal metabolite [20]. It is widespread among fungi of the Strophariaceae family, genera Gymnopilus P.karst,

Hypholoma (Fr.) P.Kumm. and Pholiota (Fr.) P.Kumm. [21]. Hispidin was originally isolated as the main compound of *Inonotus hipidus* (Bull.) P.karst from *Pistacia atlantica* tree [22,23]. We have reported in previous studies that the isolation method of hispidin, and this latter had anti-lipase activity [23-25]. It can be found in medicinal mushrooms such as *Inonotus* and *Phellinus* [26,27]. It has been reported that hispidin exhibit antiviral activity [28].

The seeds of Lepidium sativum are a rich source of alkaloids [29]. Lepidine B and E are major compounds with 1.1% and 0.88 % respectively. In a previous study, it has proved that Lepidine B and E have antimicrobial activity [30].

Curcumin (1,7-bis(4-hydroxy-3-methoxyphenyl)-1,6-heptadiene-3,5-dione) is a main phytochemical of *Curcuma longa* L. rhizome (turmeric) [31]. It has been demonstrated that curcumin has a wide range of antiviral activity against different viruses including inhibitory effect of HIV-1 and HIV-2 proteases [31-33].

This study aims to identify potent inhibitors for 2019-nCoV main protease for designing and developing drugs against the viral infection.

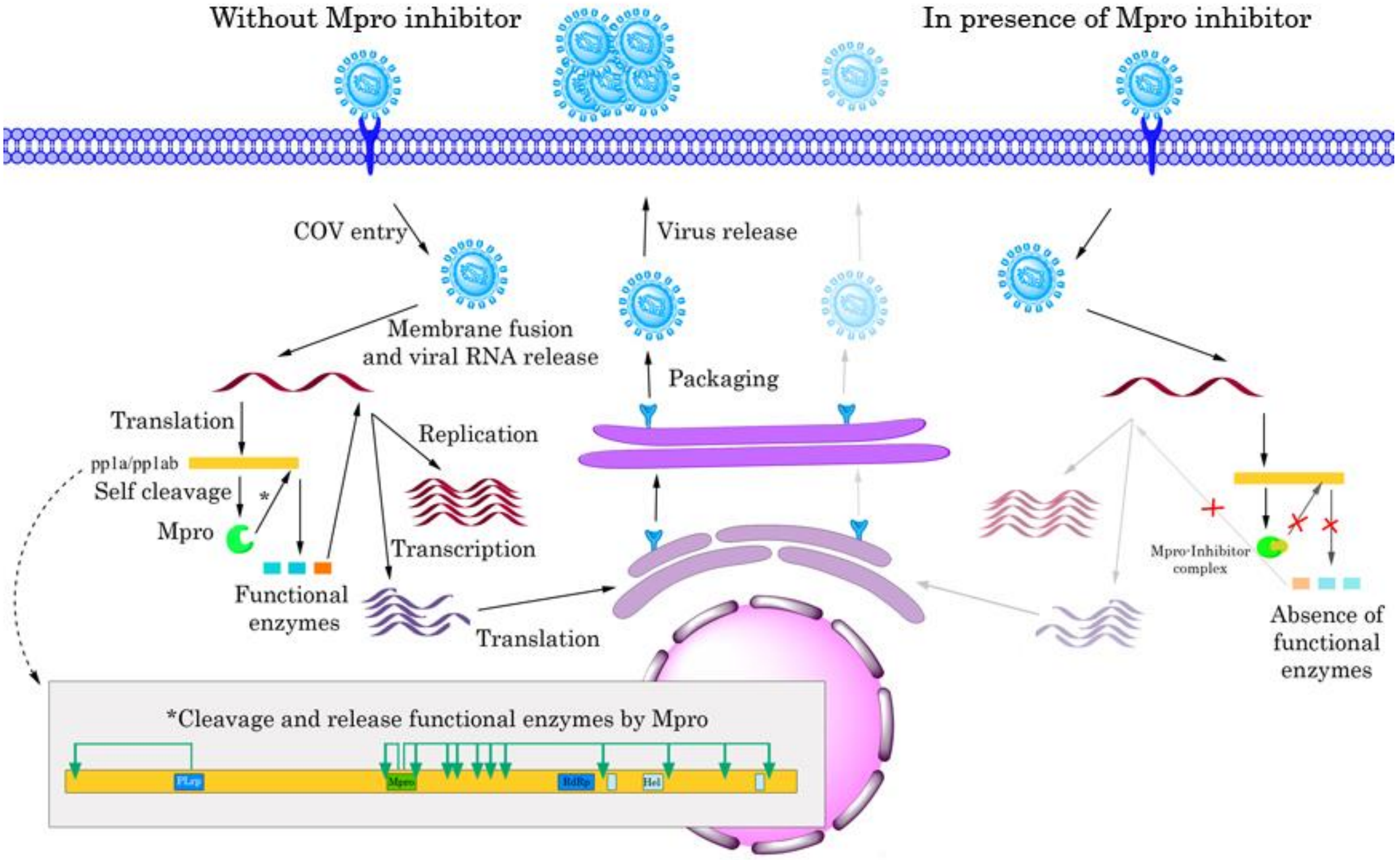

Figure. The role of Main protease in the pathogenic mechanism.

## 2. Materials and Methods

### 2.1 Sequence Alignment of Main protease

PDB and Sequence files of genomes and main protease (6lu7 and 1q2w) was download from the Protein Data Bank (PDB) [34-36], where 6lu7 is the protein sequence of 2019-nCoV and 1q2w is the protein sequence of SARS-CoV. We have conducted a brief Alignment of two structures of Mpro: 6lu7 and 1q2w, using Similarity Matrix: BLOSUM62, and BioEdit as software [37].

### 2.2 Molecular docking

The inhibitory effect of studied compounds on 3CL protease was investigated by blind and specific docking experiments. All molecules were obtained from PubChem database [38], the 2D structures of studied molecules are represented in figure1. The PDB file of the enzyme (PDB ID: 6LU7) was download from the Protein Data Bank (PDB) [36].

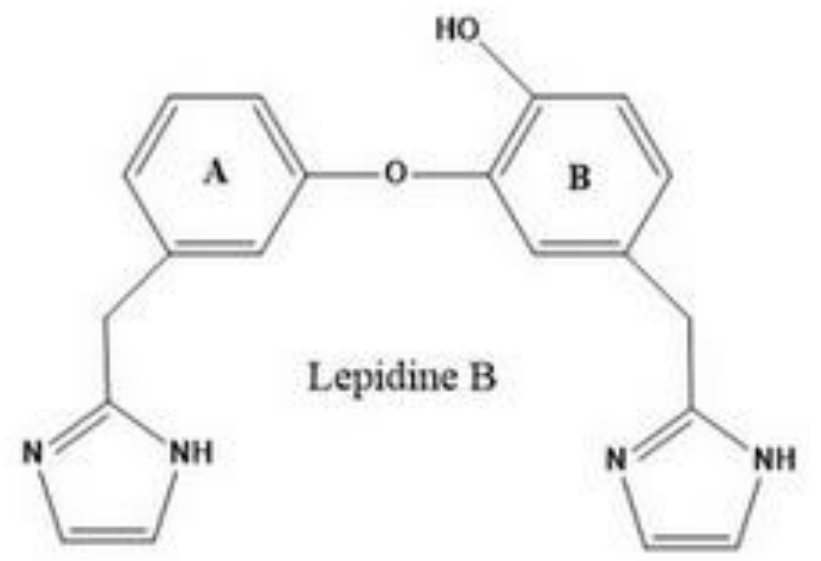

Lepidine B

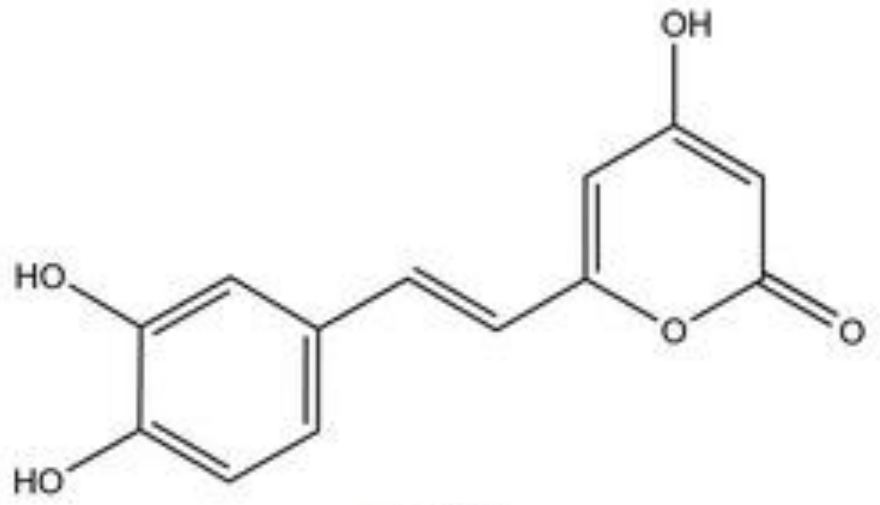

Hispidin

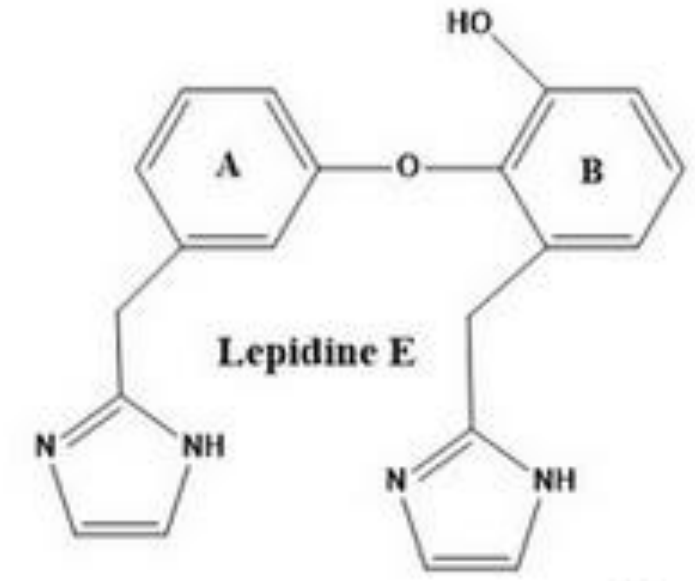

**Lepidine E**

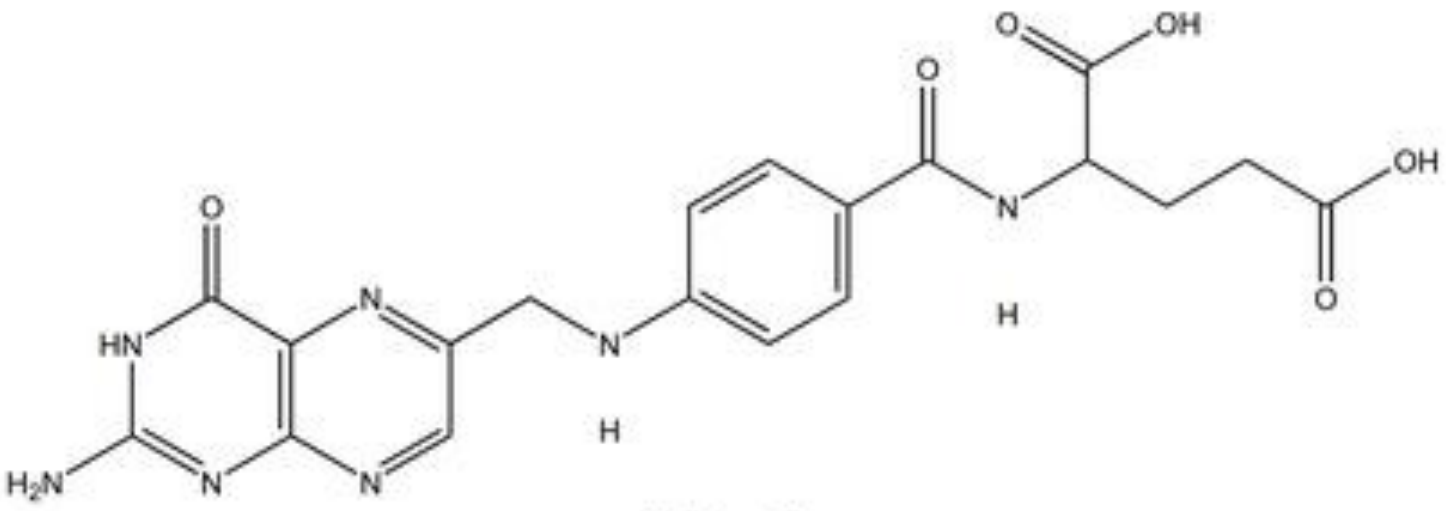

Folic acid

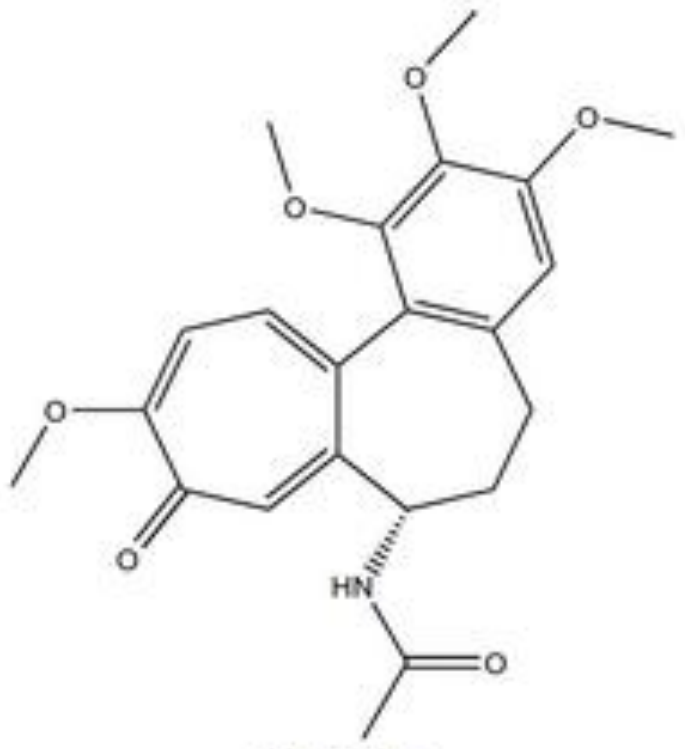

Colchicine

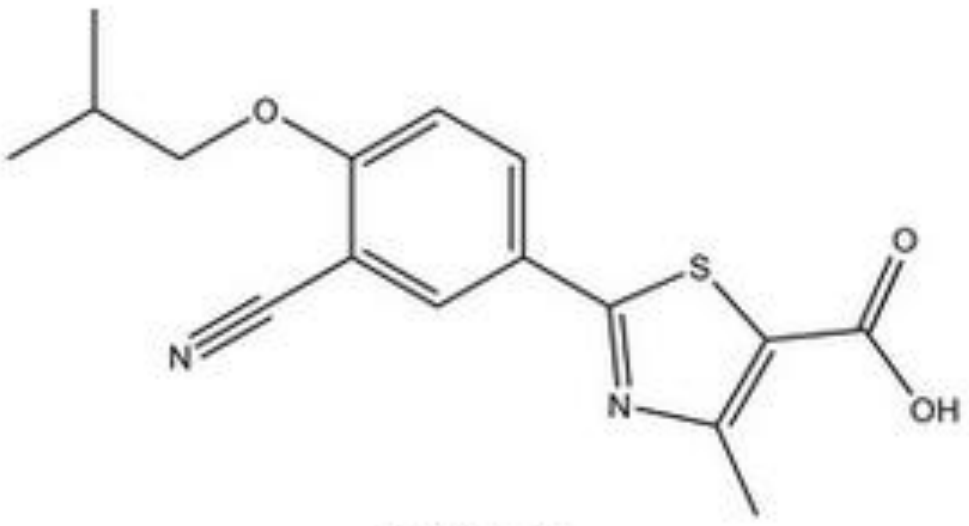

Febuxostat

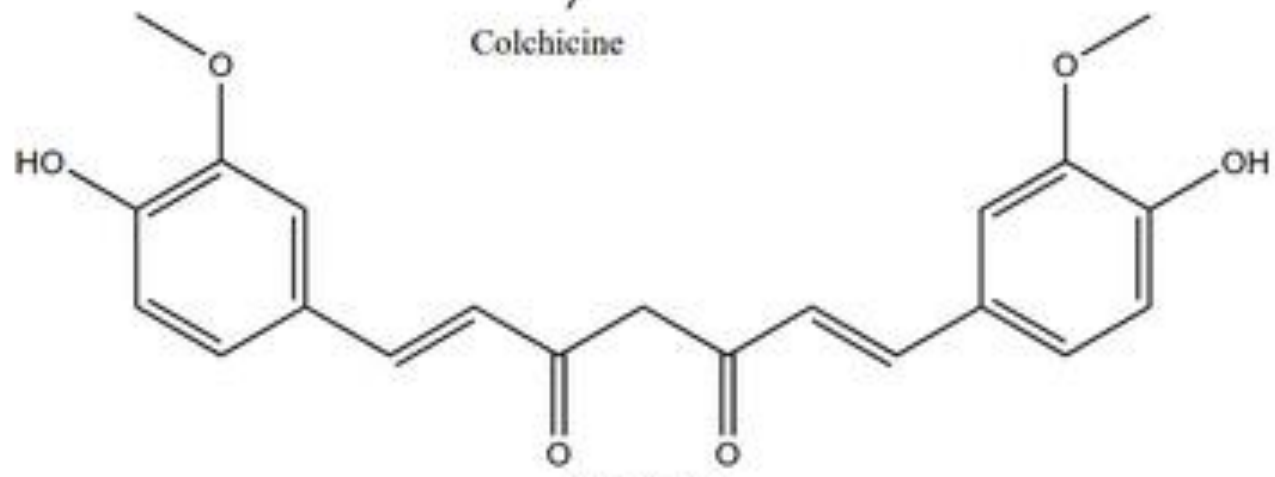

Curcumin

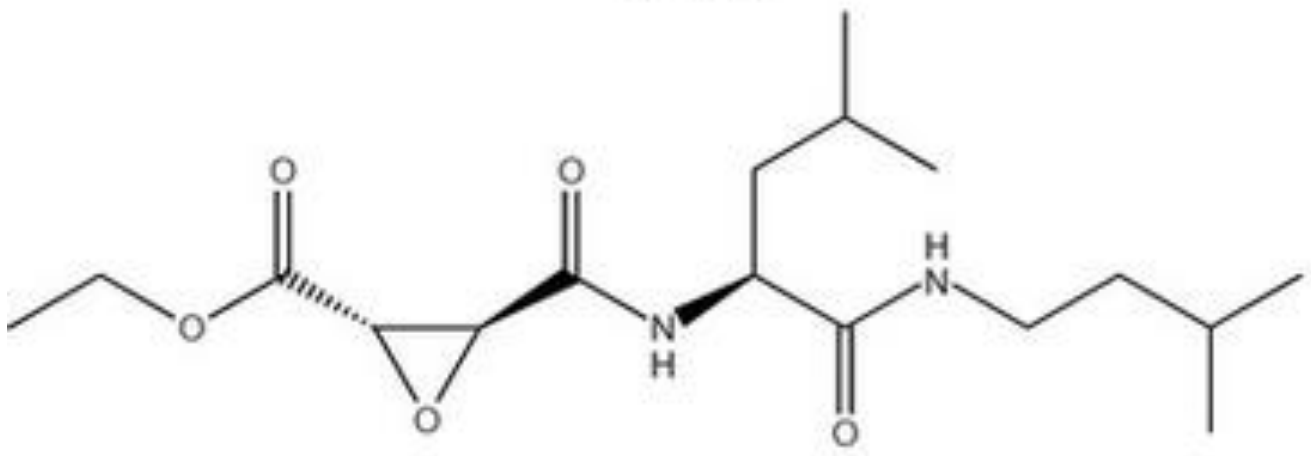

Aloxistatin

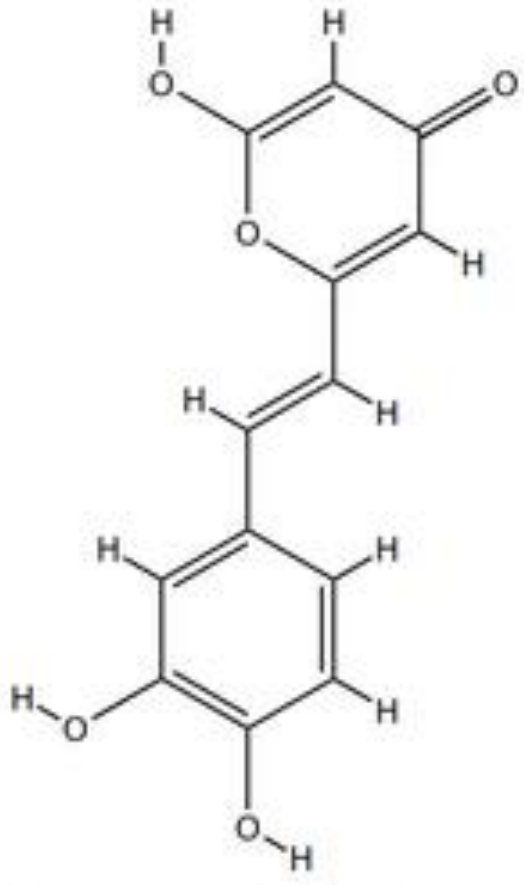

2-[(E)-2-(3,4-dihydroxyphenyl)ethenyl]-6-hydroxypyran-4-one

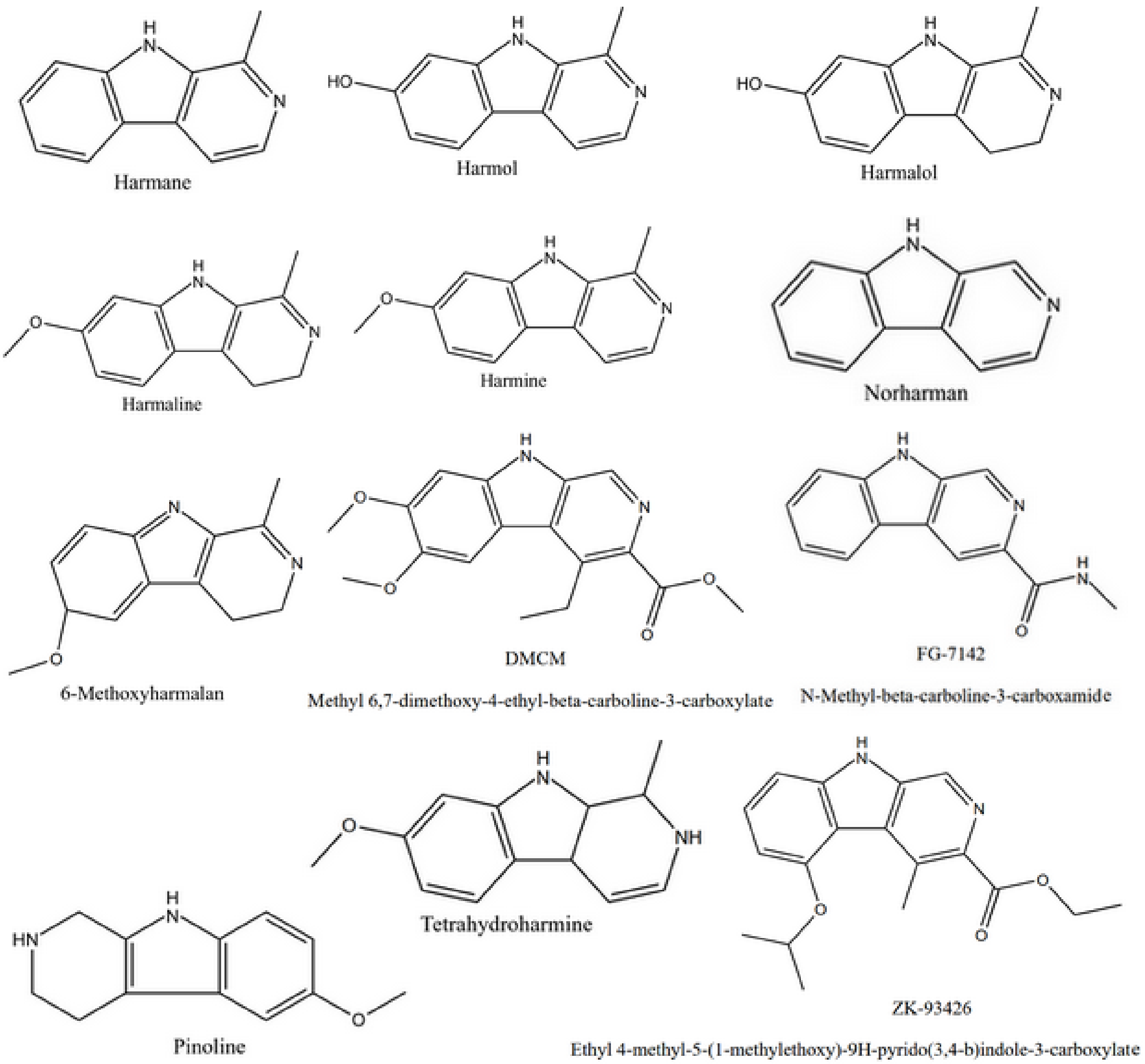

**Figure1**. 2D structure of studied molecules.

For docking, the protein must be prepared by removing all water molecules, heteroatoms, any ligands and co-crystallized solvent. Polar hydrogens and partial charges were added to the structure using Autodock tools (ADT) (version 1.5.4). Co-crystallized inhibitors have been removed. Molecular docking was performed by the AutoDock Vina program [39] in an eight CPU station. The software uses rectangular boxes for the binding site, the center of the box has been set and displayed using ADT. The box grid for 6LU7 in blind and specific docking where sets as presented in Table 1 with one Å separated grid points positioned in the middle of the active site. The default settings were used, except that the number of output conformations was set to one. The number of docking runs was set at 50 runs. The number of solutions obtained is equal to 50 conformations. All these solutions are very well handled. The "random seed" is random. The preferred conformations were those of lower binding energy within the active site. Finally, the generated docking results were directly loaded into Discovery Studio visualizer, v 4.0 [23,40,41].

**Table 1.** Docking parameters of Target drug (3CL hydrolase enzyme PDB ID: 6LU7).

| Molecular docking type | x*y*z Center of Grid Box | x*y*z Size of Grid Box |
|---|---|---|
| Blind docking | -26.283*12.602*58.961 | 52*68*68 |
| Specific docking | -12.441*12.198*67.681 | 18*22*24 |

## 3. Results

### 3.1. Sequence Alignment of Main protease

The results are presented in figure 2. 2019-nCoV main protease and SARS main protease share 94.80% sequence identity at the amino acid level. The genome of 2019-nCoV has 89% nucleotide identity with bat SARS-like-CoVZXC21 and 82% with that of human SARS-CoV [42].

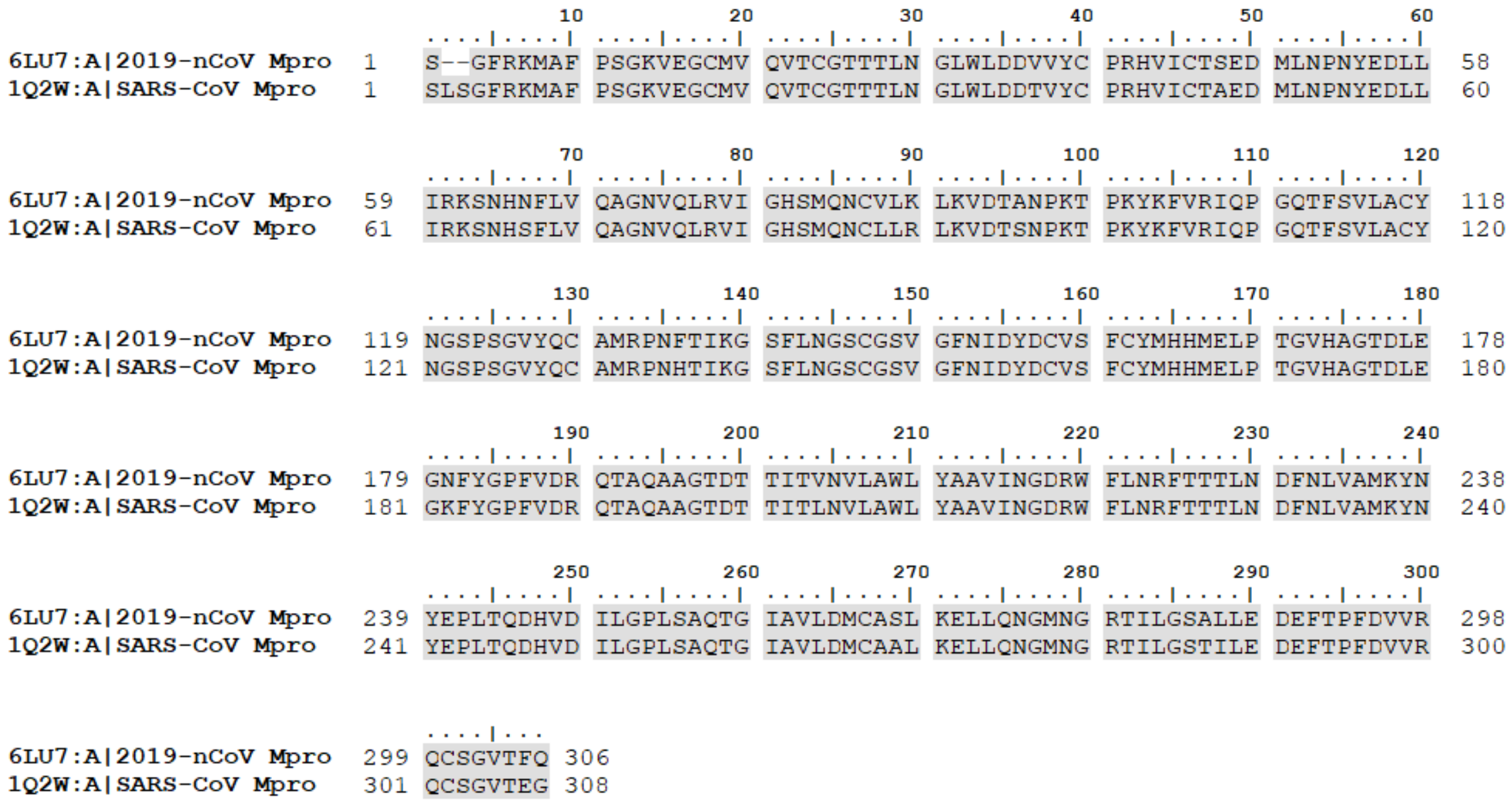

**Figure2**. Sequence alignment of 2019-nCoV main protease and SARS-CoV main protease generated in BioEdit. The white boxes represent the differences while the gray boxes represent the identity of the two amino acid sequences.

### 3.2. Molecular docking

The results of blind and specific docking for all studied molecules with the main protease (3CL hydrolase enzyme are summarized in Table 2 and 3, respectively.

**Table 2.** The results of blind docking show the main interactions of studied molecules and 2019-nCoV Protease (PDB ID: 6LU7).

| Molecule | Repeating ratio % | Affinity (kcal/mol) | Closest residues | Hydrophobic Interactions | Hydrogen bonds | Length (Å) |
|---|---|---|---|---|---|---|
| Lepidine B | 72 | -6.9 | His41, Met49, **Cys145**, **Glu166**, Pro168 | Π-Alkyl, Π- Π T-shaped, Π-Sigma | **Glu166**<br>**Glu166**<br>Gln189<br>**His163** | 3.12<br>2.83<br>2.95<br>2.64 |
| Lepidine E | 76 | -7.3 | **Thr24**, Thr25, Leu27, His41, Met49, **Cys145** | Π-Alkyl, Π- Π Stacked | **Thr26**<br>**Gly143**<br>Leu141<br>**His164** | 3.28<br>2.97<br>**1.69**<br>2.39 |
| Colchicine | 50 | -6.3 | **Asn142**, Gln189 | - | **Gly143**<br>**Asn142** | 3.27<br>2.52 |
| Folic Acid | 90 | -7.6 | Met49, **Cys15** | Π-Sigma | **Thr24**<br>Thr25<br>Ser46<br>**Cys145**<br>**His164**<br>**Phe140**<br>**Glu166**<br>**Thr26**<br>Thr45 | 3.02<br>2.85<br>2.87<br>3.00<br>2.31<br>2.62<br>2.10<br>2.48<br>**1.80** |
| Febuxostat | 38 | -6.6 | **Cys145**, Met165, Leu167, Pro168, | Alkyl, Π-Alkyl | **Gly143**<br>Ser144<br>Ser144<br>**Cys145**<br>Leu141<br>Ser144 | 3.12<br>2.85<br>3.22<br>3.06<br>2.34<br>**1.97** |
| Hispidin | **100** | -7.2 | **Cys145**, Met165 | - | Ser144<br>**Cys145**<br>Leu141<br>Ser144<br>Met165 | 3.01<br>3.30<br>**1.76**<br>2.74<br>2.96 |
| Aloxistatin | 100 | -6.3 | Met49, His41 | Alkyl, Π-Sigma | **His164**<br>His41<br>**Gly143** | 2.97<br>3.28<br>3.30 |
| Curcumin | 80 | -6.5 | Met49, Tyr45, Met165 | Π-Alkyl | Ser144<br>**Cys145**<br>**Cys145** | 3.26<br>3.32<br>3.34 |
| DHPEHP | 100 | -7.2 | **Cys145**, Met165 | - | Arg188<br>**His163**<br>Ser144<br>Ser144<br>**Gly143** | 2.08<br>2.14<br>2.39<br>2.93<br>3.19 |
| DCMC | 86 | -6.2 | His41, Tyr54, Met165, **Glu166**, Gln189 | Π-Alkyl, Π-Sigma | Arg188 | 2.28 |
| FG-7142 | 84 | -6.8 | **Cys145**, Met49 | Π-Alkyl | **Glu166**<br>**Asn142** | 3.23<br>2.65 |
| Harmaline | 100 | -6.3 | His41, Met49, Met165, | Alkyl, Π-Alkyl | - | - |
| Harmalol | 38 | -5.9 | **Gly143**, **Cys145** | Π-Alkyl | **Gly143**<br>**Cys145**<br>**His163** | 3.28<br>3.62<br>3.10 |
| Harmane | 98 | -6.1 | His41, **Cys145**, Met165 | Alkyl, Π-Alkyl, Π- Π T-shaped | - | - |
| Harmine | 100 | -6.3 | His41, **Cys145**, Met165 | Alkyl, Π-Alkyl, Π- Π T-shaped | - | - |
| Harmol | 86 | -6.3 | His41, Met165 | Π-Alkyl | - | - |
| 6-Methoxyharmalan | 98 | -6.3 | His41, Met49, Met165 | Alkyl, Π-Alkyl | - | - |
| Norharman | 70 | -5.7 | His41, Met165 | Π-Alkyl, Π- Π T-shaped | - | - |
| Pinoline | 100 | -6.0 | His41, Met165, Gln189, | Π-Alkyl, Π-Sigma, Π-Π T-shaped | Arg188 | 2.16 |
| Tetrahydroharmine | 78 | -6.2 | His41, Met165, Gln189 | Π-Alkyl, Π-Sigma | Arg188 | 2.35 |
| Zk-93426 | 56 | -6.8 | His41, **Asn142**, **Cys145**, Met49, Met165, **Glu166** | Alkyl, Π-Alkyl, Π-Sigma | **Gly143**<br>Ser144<br>**Cys145** | 3.00<br>3.15<br>3.16 |

**Table 3.** The results of specific docking show the main interactions of studied molecules and 2019-nCoV Protease (PDB ID:6LU7)

| Molecule | Repeating ratio % | Affinity (kcal/mol) | Closest residues | Hydrophobic Interactions | Hydrogen bonds | Length (Å) |
|---|---|---|---|---|---|---|
| Lepidine B | 90 | -7.1 | His41, Met49, **Cys145**, Met165, Gln189 | Π-Alkyl, Π- Π Stacked | **Glu166**<br>**His164**<br>**His163** | 3.09<br>2.70<br>2.90 |
| Lepidine E | 70 | -7.8 | Thr25, Leu27, His41, Met49, **Cys145** | Π-Alkyl, Π-Sigma, Π-Π Stacked | **Gly143**<br>**Cys145**<br>**Thr24**<br>Leu141<br>**His164** | 2.88<br>3.38<br>2.32<br>**1.74**<br>2.38 |
| Colchicine | 100 | -6.3 | **Glu166** | - | - | - |
| Folic acid | 52 | -7.4 | Leu141, **Asn142**, **Gly143**, **Cys145**, **Glu166**, Gln189 | Π-Alkyl | **Thr26**<br>**Glu166**<br>Thr190<br>Ser144<br>**Cys145**<br>Gln192 | 2.12<br>2.13<br>**1.74**<br>3.18<br>3.13<br>2.88 |
| Febuxostat | 100 | -6.9 | **Cys145**, **His163**, Met165, **Glu166**, Pro168 | Alkyl, Π-Alkyl | **Gly143**<br>Ser144<br>Ser144<br>**Cys145** | 2.97<br>2.95<br>3.11<br>3.08 |
| Hispidin | 100 | -7.2 | **Cys145**, Met165 | - | Ser144<br>**Cys145**<br>Leu141<br>Leu141<br>Ser144<br>Met165 | 3.01<br>3.31<br>2.83<br>**1.77**<br>2.86<br>2.93 |
| Aloxistatin | 100 | -6.5 | His41**, His163** | Π-Alkyl | His41<br>**Gly143**<br>Ser144<br>**Glu166** | 3.17<br>2.95<br>3.37<br>3.27 |
| Curcumin | 90 | -7.5 | His41**,** Asn142**, Cys145,** Met165 | Π-Alkyl, Π- Π T-shaped | Gly143<br>Gly143<br>Ser144<br>Ser144 | 3.16<br>3.38<br>2.96<br>3.25 |
| DHPEHP | 100 | -7.2 | **Cys145,** Met165**,** Gln189 | - | Ser144<br>Leu141<br>Ser144 | 3.20<br>2.01<br>2.68 |
| DCMC | 76 | -6.9 | Met49, Tyr54, **Cys145** | Π-Alkyl | Ser144 | 3.19 |
| FG-7142 | 100 | -6.8 | **Cys145**, Met49, **Phe140**, **Cys145** | Π-Alkyl | **Glu166**<br>**Asn142** | 3.21<br>2.63 |
| Harmaline | 100 | -6.3 | His41, Met49, Met165 | Alkyl, Π-Alkyl | - | - |
| Harmalol | 100 | -6.1 | **Gly143**, **Cys145** | Π-Alkyl | **Gly143**<br>**Cys145**<br>**His163** | 2.98<br>3.63<br>3.02 |
| Harmane | 100 | -6.1 | His41, **Cys145**, Met165 | Alkyl, Π-Alkyl, Π- Π T-shaped | - | - |
| Harmine | 100 | -6.3 | His41, **Cys145**, Met165 | Alkyl, Π-Alkyl, Π- Π T-shaped | - | - |
| Harmol | 100 | -6.3 | His41, Met165 | Π-Alkyl, Π- Π T-shaped, Π-Sigma | - | - |
| 6-Methoxyharmalan | 100 | -6.4 | His41, Met49, Met165 | Alkyl, Π-Alkyl | - | - |
| Norharman | 100 | -5.7 | His41, Met165 | Π-Alkyl, Π- Π T-shaped | - | - |
| Pinoline | 100 | -5.9 | His41, Met165, Gln189 | Π-Alkyl, Π- Π T-shaped, Π-Sigma | Arg188 | 2.14 |
| Tetrahydroharmine | 100 | -6.3 | His41, Gln189, Met165, **Glu166** | Π-Alkyl, Π-Sigma | Arg188 | 2.37 |
| Zk-93426 | 82 | -6.9 | **Cys145**, Met165, **Glu166** | Alkyl, Π-Alkyl | **Glu166** | 2.73 |

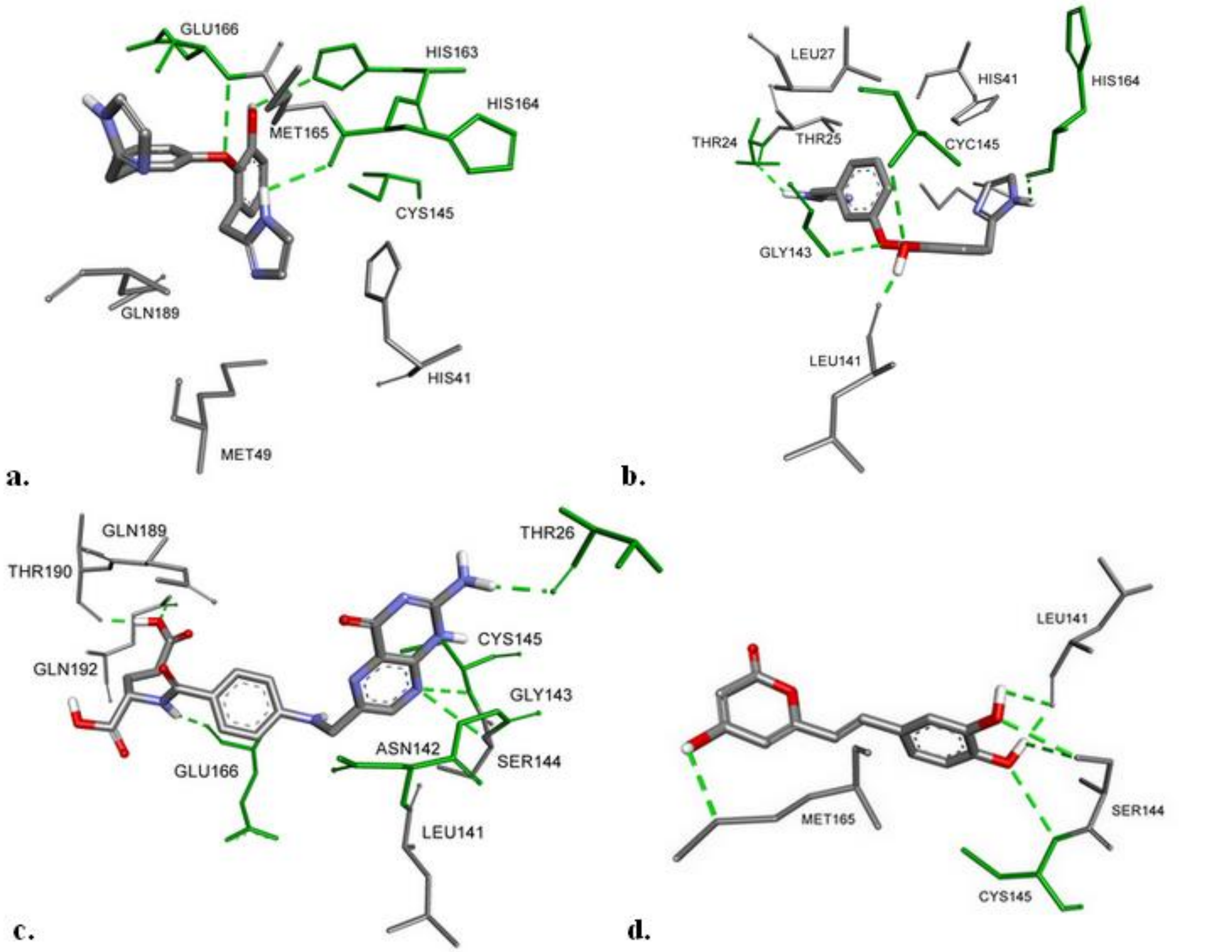

**Figure 3.** Binding mode of lepidine B (a), lepidine E (b), folic acid (c), and hispidin (d) at the active site of with 6LU7. Ligand are illustrated in gray, rendered as sticks. Different amino acids of the Active site were colored in green.

## 4. Discussion

Chymotrypsin-like protease (3CLpro) called also the main protease (Mpro) [43] is suggested to be a potential drug target to combat 2019-nCoV. With a view to identify inhibitors of 2019-nCoV Mpro, twenty-one drugs were selected based on their structure and the conformation of Mpro active site. As presented in table 2 and 3, the results show that three molecules (hispidin, lepidine E and folic acid) could bind tightly with the enzyme. His41, Cys145, and Glu166 are important residues in the substrate-binding subsite S1 for the proteolytic activity [44], the involvement of these residues in forming hydrogen bonds is important for the inhibitory effect of the Mpro. The molecular docking study demonstrated that most of studied molecules bind with one or more of these residues.

The best molecule is hispidin. The blind docking shows that 100% of poses are in the active site, those poses share the same position and interactions with 2019-nCoV protease residues (Fig.3 (d)). The predicate binding affinity to the 2019-nCoV protease is -7.2 Kcal/mol. Hispidin forms a strong hydrogen bond network with nCoV-2019 protease, six hydrogen bonds have been observed in both blind and specific docking. The strongest hydrogen bond (1.77Å) is formed by the hydroxyl group of the benzene ring of hispidin and CO of Leu141. Hydroxyl groups of hispidin and residues Leu141, Ser144, Cys145 and Met165 of 2019-nCoV protease form the other hydrogen bonds. No hydrophobic interactions have been recorded.

The second molecule is lepidine E, 76% of poses are in the active site. We ran specific docking to study the binding mode of lepidine E in the active site (Fig.3 (b)). Lepidine E bind tightly with the 2019-nCoV protease residues with a binding affinity -7.8 kcal/mol.

A very strong and short hydrogen bond has been observed between hydroxyl groups of lepidine E and Leu141 with length of 1.64Å and 1.74Å in blind docking and specific docking respectively. Which mean it can form a covalent bond that may lead to irreversible inhibition. The other atoms of lepidine E forms hydrogen bond with 2019-nCoV protease residues, whereas, NH of both imidazole rings bind to Oxygen atom (of main chain) of Thr24and His164 (with interatomic distances "ID" of 2.32 and 2.38Å). Oxygen atom and hydroxyl group interact with NH main group of Gly143 and Cys145 forming hydrogen bond with length of 2.88 and 3.38Å, respectively. This binding supported by hydrophobic interactions with Thr24, Leu27 and His41 by means Π-sigma, Π-alkyl, and Π-Π-stacked, respectively. It is noticed that lepidine B bind to 2019-nCoV protease differently, the hydrogen bonds are of length 2.7, 2.9 and 3.09Å with His164, His163, and Glu166 respectively (Fig.3 (a)).

The third best molecule is folic acid. The blind docking shows that 90% of poses are in the active site with a binding affinity of -7.6 kcal/mol. Different positions have been recorded in specific docking. Despite this, we observe that all positions contain hydrogen bonds, from 5 to 11 hydrogen bonds in one position. In the most frequented pose, we record 6 hydrogen bonds (Fig.3 (c)). The strongest hydrogen bond (1.74Å) between this molecule and the protease is the carboxyl group of the molecule and the oxygen atom of Thr190 in the main chain. The carboxyl group of the molecule also forms hydrogen bond with N in the side-chain of the residue Gln192 (2.88Å). The additional hydrogen bonds are as follows: $NH_2$ of the molecule with CO of Thr26, NH with CO of Glu166, and N of the pteridine ring of folic acid bind to NH main group of Ser144 and Cys145. These hydrogen bonds lead to a strong binding of 2019-nCoV protease binding site.

In PDB file of crystal structure of main protease with the inhibitor, we observed that the inhibitor was bind weaker than our molecules. It forms nine hydrogen bonds with 2019n-CoV main protease's residues: Phe140, Gly143, His163, His164, Glu166, Gln189, and Thr190. All hydrogen bond's length was higher than 2Å. The shortest hydrogen bond was between His163 and PJE5 of the inhibitor with length of 2.43Å.

Curcumin shows also a strong binding affinity with -7.5 kcal/mol. In most repeated pose (90%), four hydrogen bonds have been formed with Gly143 and Ser144 (2.96-3.38 Å). It has been found that curcumin has an inhibitory effect against of HIV-1 and HIV-2 proteases [31-33].

Colchicine, febuxostat, aloxistatin, DHPEHP, and β-carboline derivatives bind strongly in the binding site of 2019-nCoV protease. β-carboline was stabilized in the active site mostly by hydrophobic interactions. The binding affinity was -6.9 to -5.7 Kcal/mol except DHPEHP where the binding affinity was -7.2, but the length of the hydrogen bonds was always bigger than 2Å. No hydrogen bond was formed with colchicine, Harmaline, Harmane, Harmine, Harmol, Methoxy, and Norharman.

Since publishing the crystalized structure of 2019-nCoV main protease, many preprint papers have performed *insilico* studies for identify an inhibitor for it. A number of 1903 drugs were tested as the Mpro inhibitors in a preprint manuscript, among them, Nelfinavir was proposed to be potential inhibitor against 2019-nCoV [45]. In another study, it was found by screening 8,000 clinical drug libraries that 4 small molecular drugs have high binding capacity with the main protease [46]. No binding manner or hydrogen bonds formation with active site residues has been discussed in these studies. It was declared that 10 commercial medicines may form hydrogen bonds with key residues of 2019-nCoV main protease [47]. According to articles published in Nature Communications and Nature reviews drug discovery (Sheahan *et al,* 2020 and Guangdi *et al,* 2020) [48,49], currently, only four drugs (Ritonavir, Darunavir, cobicistat, and ASC09F) are considered for treatment of 2019-nCoV as 3CLpro inhibitors. Through this study, we recommend to enter these molecules (Hispidin, Lepidine E and Folic acid) in the therapy of this disease especially that the studied molecules are known and all the necessary clinical experiments are already done, for the natural compounds, they are used in folk medicine to treat several diseases.

# 1 Conclusions

We have found that hispidin, lepidine E, and folic acid display a strong inhibitory activity on 3CL enzyme, which prevents spreading the infection by stopping the virus's cycle life. Based on this study's results and other previous studies, we propose a therapeutic strategy to prevent and treat the virus infection. The potent metabolites hispidin and folic acid that inhibits the main protease of 2019-nCoV might be an effective strategy to treat 2019 novel coronavirus infected individuals. Those natural molecules also might become drug candidates as anti-CoViD-19 drug.

**LIST OF ABBREVIATIONS**

**2019-nCoV**: 2019 novel coronavirus (the virus)

**3CL hydrolase**: 3-chymotrypsin-like hydrolase

**CoViD-19:** coronavirus disease 2019 (the illness)

**MERS:** Middle East respiratory syndrome

**Mpro:** Main protease

**SARS:** Severe acute respiratory syndrome

### HUMAN AND ANIMAL RIGHTS

No human or animal experiments have been performed.

### ACKNOWLEDGMENTS

The authors thank Zihe Rao and Haitao Yang's research team at Shanghai Tech University for providing the crystal structure of 2019-nCoV 3CL hydrolase (PDB ID 6lu7).